\documentclass[12pt,letterpaper]{amsart}
\usepackage{setspace}
\usepackage{setspace}

\usepackage{amsmath}%
\usepackage{amsfonts}%
\usepackage{amssymb}%
\usepackage[foot]{amsaddr}%

\usepackage{graphicx}

\usepackage[numbers]{natbib}
\newlength{\fw}
\begin{document}

\newcommand{\W}{16cm}

\title[Vector separation of particles and cells]{Vector separation of particles and cells using an array of slanted open cavities}

\author{Jorge A. Bernate$^{\ast}$\textit{$^{a}$} }
\author{Chengxun Liu\textit{$^{b}$} }
\author{Liesbet Lagae\textit{$^{b,c}$} }
\author{Konstantinos Konstantopoulos\textit{$^{a}$}}  
\author{German Drazer$^\ddag$\textit{$^{a}$}}
\address{\textit{$^{a}$~Department of Chemical and Biomolecular Engineering, Johns Hopkins University, Baltimore MD, 21218}}
\address{\textit{$^{b}$~Interuniversity Microelectronics Center (IMEC), Kapeldreef 75, Heverlee 3001}}
\address{\textit{$^{b}$~Katholieke Universiteit Leuven, Leuven 3000, Belgium}}
\address{\textit{${\ast}$~ Current address: Department of Chemical Engineering, Stanford University, Stanford, CA 94305, USA}}
\address{\textit{${\ddag}$~ Current address: Mechanical and Aerospace Engineering Department, Rutgers, The State University of New Jersey, Piscataway, NJ 08854, USA}}


\begin{abstract}
We present a microfluidic platform for the continuous separation of suspended particles based on their size and settling velocity. 
The separation method takes advantage of the flow field in the vicinity and inside a parallel array of slanted open cavities. These cavities induce flow along them,
which deflects the suspended particles to a different degree depending on the extent to which they penetrate into the cavities. 
The cumulative deflection in the periodic array ultimately leads to vector chromatography, with the different species in the sample moving in different directions.  
We demonstrate density and size based separation over a range of flow rates by separating polystyrene and silica particles and show that purities nearing 100\% 
can be achieved for multicomponent mixtures. We also demonstrate the potential of the platform to separate biological cells by fractionating different blood components.
We discuss the presence of two regimes, which can be distinguished depending on the ratio between the settling velocity 
and the velocity of the particles across the open cavities. The proposed platform could also integrate additional separative force fields 
in the direction normal to the plane of the cavities  to fractionate specific mixtures based on the distinguishing properties of 
the component species.
\end{abstract}

\maketitle

\section{Introduction}
 
Micro- and nano-fluidic separation stages are crucial components of Micro Total Analysis Systems ($\mu$TAS) and other integrated systems in lab-on-a-chip platforms.\cite{Blow:2007,Blow:2009} 
In fact, many applications are ultimately a separation per se, such as in diagnostic systems where the goal is to detect/isolate a target of interest. 
In other applications, a separation stage is necessary to enrich a particular species present in a multicomponent sample for downstream processing. 
Several strategies have been developed to separate particles, biological and otherwise. 
Depending on the operation mode, these strategies have been broadly classified as batch or continuous methods and, depending on the use of external fields to drive separation, 
they are considered passive or active techniques. 
Continuous methods, are generally less labor intensive and provide higher throughputs.\cite{Pamme:2007,Kersaudy-Kerhoas:2008,Kulrattanarak:2008,Lenshof:2010} 
Passive methods, including pinch flow fractionation,\cite{Yamada:2004} hydrodynamic filtration,\cite{Yamada:2005} inertial focusing,\cite{DiCarlo:2007,DiCarlo:2008,Park:2009,Sim:2011} 
and deterministic lateral displacement,\cite{Huang:2004,Inglis:2006} are easier to operate and more portable than active methods which usually require bulky external components. 
In one approach to passive separation, geometrical features are fabricated into the device to create flow fields that drive
different particles in distinctive trajectories. 
Separation systems based on ridged surfaces, for example, rely on the recirculation flow that ensues in the main channel, when the ridges are not perpendicular to the channel.
In general, big and small particles are carried by oppositely bound streams of the recirculation flow and can be separated.\cite{StroockScience:2002,StroockAC:2002,Choi:2007,Choi:2008,Chen:2008,Hsu:2008} 
In these systems, the magnitude of the recirculating velocity is enhanced by using devices in which the depth of the grooves is comparable to the dimensions of 
the channel.\cite{StroockScience:2002,StroockAC:2002,Choi:2007,Choi:2008,Chen:2008,Hsu:2008}. 
Passive methods, however, are usually less versatile than active ones, 
which can use specific fields to fractionate a mixture based on the differentiating properties of the different species, 
e.g., electric charge and impedance (electrophoresis\cite{Krivankova:2005} and dielectrophoresis~\cite{Rousselet:1994,Kralj:2006}), refractive index (optical sorting~\cite{Gluckstad:2004,Xiao:2010}), compressibility (acoustophoresis~\cite{Petersson:2007}), and magnetic susceptibility (magnetophoresis~\cite{Inglis:2004,Pamme:2006,Liu:2009}). 
Among both passive and active strategies, continuous two-dimensional techniques, also termed vector separation or vector chromatography (VC), are particularly promising. 
In VC, the different species in a sample fan out in different directions, and thus separate laterally, allowing their continuous collection 
with high resolution.\cite{Giddings:1984,Tia:2009,Dorfman:2001,Arata:2009,Mao:2011,Bernate:2011,Bernate:2012} 

Here, we present a microfluidic device that takes advantage of the characteristics of the flow field in the vicinity and inside slanted open cavities patterned at the bottom surface of straight channels. We suppress the recirculation that ensues in confining geometries using devices with dimensions much larger than the dimensions of the ridges and the cavities that they create, and exploit the reorientation of the flow inside the open the cavities
to passively and continuously fractionate, via vector chromatography, suspended particles. In this system, separation can occur based on particle size, but also based on 
the settling velocity of the different particles, which represents an extra degree in versatility with respect to previous separation devices in which VC is based on a single property. 
Note that the settling velocity of a particle depends on its shape, size, and buoyant density. 
Furthermore, it is straightforward to use alternative (additional) force fields to play (enhance) the role of gravity in the present device, 
including electric, dielectrophoretic, and magnetic, which opens the door to a broad range of applications.

We shall show that two regimes can be distinguished depending on the ratio between the settling velocity of the particles  
and their average velocity in the direction across the cavities. Alternatively, we can consider the ratio
between the characteristic time for a particle to settle inside the cavities and their average time to cross them. 
At low flow rates, when the particles have time to sediment as they are advected by the fluid flow across the cavities, 
heavier ones settle deeper and are deflected for a longer time than lighter particles. 
On the other hand, if sedimentation is negligible, particles can be separated based on their size, 
with smaller particles carried deeper into the cavities by the flow field and therefore deflecting more than larger particles. 
In the first part of this work, we gain insight into the separation mechanism by numerically solving the particle-free flow. 
We then present experiments separating beads based on their size and settling velocity. 
We also demonstrate the potential of this method to separate cells by showing the individual deflection experienced by
different blood components.

\section{Materials and methods}
\subsection{Device design and fabrication}

\begin{figure}[htbp]
\includegraphics[width=\columnwidth]{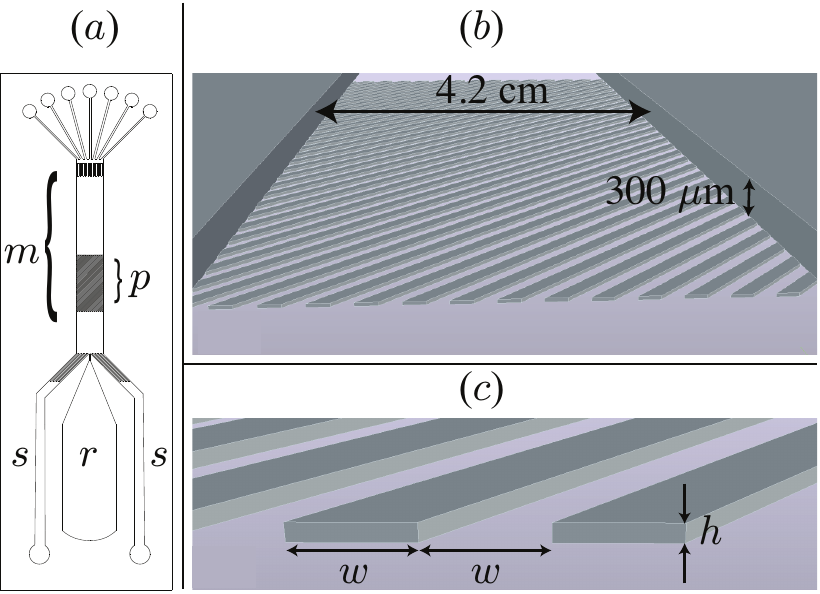}	
\caption{$(a)$ Layout of the device showing the side channels ($s$) use to flow focus particles and cells from the settling reservoir ($r$) into the main channel (m) and towards the patterned region ($p$). The ridges are oriented at an angle of $45^{\circ}$ with respect to the main channel. $(b)$ and $(c)$ show perspective views of the patterned region. The dimensions of the main channel are much larger than the width $(w)$ and height $(h)$ of the ridges ($h = 10$ $\mu$m, $w = 50$ $\mu$m and $h = 24$ $\mu$m, $w = 100$ $\mu$m).}
\label{Fig:Device}
\end{figure}

Fluidic channels were made of PDMS using a standard molding and casting procedure. 
In short, a mold was made on a silicon wafer using standard photolithography. 
Two layers of the negative photoresist SU-8 3050 (Microchem) were used to give features 300 microns tall. 
A PDMS negative replica was then cast using this mold.
The patterned surface consists of a microscope glass slide (1'x3''x1mm) that is patterned with a periodic array of protruding  rectangular ridges. 
The ridges were fabricated using standard photolithography and the photoresist SU-8 3025 (Microchem).
A sealed microfluidic device is then obtained by irreversibly bonding the PDMS fluidic layer with the patterned glass slide after activating both surfaces with oxygen plasma. 
Figure~\ref{Fig:Device}$(a)$ shows a top-view schematic of the device; sheath flows from the side channels ($s$) allowed to flow focus particles and cells from the settling reservoir ($r$) into the main channel $(m)$ and towards the patterned region ($p$) (also shown in perspective view in Fig.~\ref{Fig:Device}$(b)$). Experiments were performed with two different devices 
with the same dimensions for the fluidic layer and patterned region but with ridges of different height and width (see Fig.~\ref{Fig:Device}$(c)$ for a schematic of the ridges)
The dimensions of the {\it small} ({\it large}) ridges are $h = 10$ $\mu$m and $w = 50$ $\mu$m  ($h = 24$ $\mu$m, $w = 100$ $\mu$m).

\subsection{Samples preparation}
4.32 $\mu$m silica (SiO$_2$) particles, 4.31 and 10.11 $\mu$m polystyrene (PS) 
(Bangs Laboratories Inc., Fishers, IN), 10.11 and 20.9 $\mu$m SiO$_2$ particles  (Microspheres-Nanospheres, Corpuscular, Cold
Springs, NY), and 20.9 $\mu$m PS particles (Duke scientific Corp., Fremont, CA) were used in the experiments 
(the specified dimension corresponds to the nominal diameter of the spherical particles). 
The density of SiO$_2$ particles is $\rho$ = 1.96 g/mL, and that of PS particles is $\rho$ = 1.06 g/mL.
Dispersions were prepared in deionized (DI) water with a concentration of approximately 10$^6$  mL$^{-1}$ for each particle used in a given experiment.
Blood samples were collected by venipuncture into 5 mL ACD (anticoagulant citrate dextrose) vacutainer tubes (BD Franklin Lakes, NJ). 
White blood cells (WBCs)  were concentrated as follows. 
The two vacutainer tubes were centrifuged at 1000 rpm for 5 mins  immediately after blood collection. 
The upper layer and buffy coat from each vacutainer tube was transferred to different centrifuge tubes (1 tube per vacutainer) and 
1\% (w/w) bovine serum albumin (BSA) in phosphate buffer saline (DPBS, Gibco, life technologies) was added to a volume of 3 mL.  
The two vacutainer tubes were centrifuged once more at 1000 rpm for 5 mins. 
The upper layer and buffy coat from each vacutainer tube were again transferred to the same centrifuge tubes used before. 
1\% BSA in DPBS was added to each of the centrifuge tubes to a volume of 5 mL before centrifugation at 1000 rpm for 5 mins. 
After removing the supernatant, the pellets were resuspended in 0.5\% BSA in DBPS  to a total volume of 1 mL. 
10 $\mu$L of whole blood were added to the concentrated WBCs suspension.

\subsection{Device pretreatment, sample injection, microscopy and data collection}
\label{subsec:methods}

The fluids (DI water and 0.5\% BSA in DPBS for the particles and blood experiments, respectively) were pumped through the device using a pressure-driven flow system. 
Each inlet was pressurized using a 0-5 psi regulator and was equipped with a switching valve (IDEX health \& science) that was used to route fluid to the device, 
either from the pressurized containers or from a syringe. The syringes were used to prime and clean the device between experiments. 
The device was first flushed by manually injecting 5 mL of  5\% Alconox detergent in DI water through each of the three inlets. 
If necessary, the device was sonicated while being flushed to remove adhered material. 
For the particle experiments, the device was simply flushed with the same volume of DI water in the same way as when flushing with the Alconox solution. 
For the blood experiments, the device was similarly flushed with water, then with DPBS, and finally with 2.5 \% BSA in DPBS. 
The device was then left blocking overnight at 4 $^o$C. Prior to the injection of a blood sample, the device was flushed with 0.5 \% BSA. 
After priming and pretreatment, the device was mounted on an upright transmitted light microscope for the particle experiments and on an inverted phase contrast microscope for the blood experiments. 
The samples were manually injected into the settling reservoir immediately after preparation. 
The particles and the blood cells were allowed to sediment to the bottom surface of the device before 
flowing them towards the patterned region. 
In the case of blood, cells were flow focused after no more red blood cells (RBCs) and WBCs were observed to arrive at the bottom surface of the settling reservoir 
(platelets were still distributed over the height of the device).  
WBCs, RBCs, and platelets could be readily distinguished under the phase contrast microscope without the need of fluorescence dyes. 
Platelets are clearly smaller than the rest of the blood cells. 
WBCs appear round and do not change shape or brightness considerably as they flow. 
In contrast, stationary RBCs appear dark in the center and bright in the surrounding area. 
Moreover, RBCs exhibit different modes of motion and ``blink'' as they flow at moderate speeds. 
At high velocities, however, RBCs align with the flow and appear as bright ellipsoids of constant intensity (see ESI Movies 1 and 2 for representative videos of flowing RBCs, WBCs, and platelets). 
Videos of the particles and cells were captured in the middle of the cross-section of the device 
in the plain region as they approached the patterned area and about 1 mm into the patterned area to measure 
the approach velocity and the deflection angle. We define the deflection angle $\Delta \theta$  as the difference between the approach 
angle and the migration angle of the particles or cells on the patterned region (see ESI for details).

\section{Separation principle}	
\begin{figure}[htbp]
	\begin{center}
		\begin{tabular}{c|c}
			$(a)$ & $(b)$ \\ 
			 \parbox[][][c]{8cm}{
	          		\includegraphics[width=8cm]{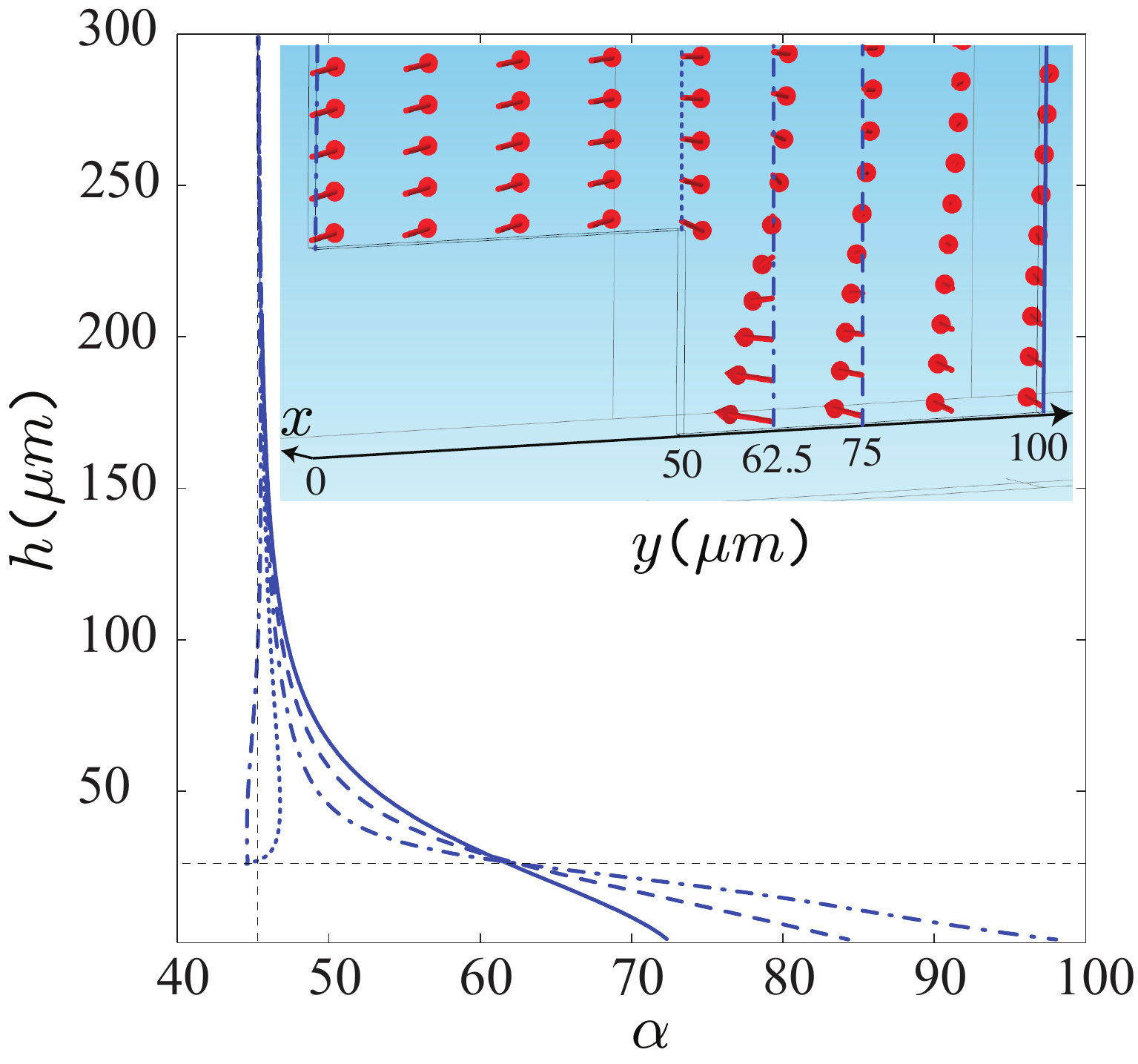}} &
	     \parbox[][][c]{8cm}{
	               \includegraphics[width=8cm]{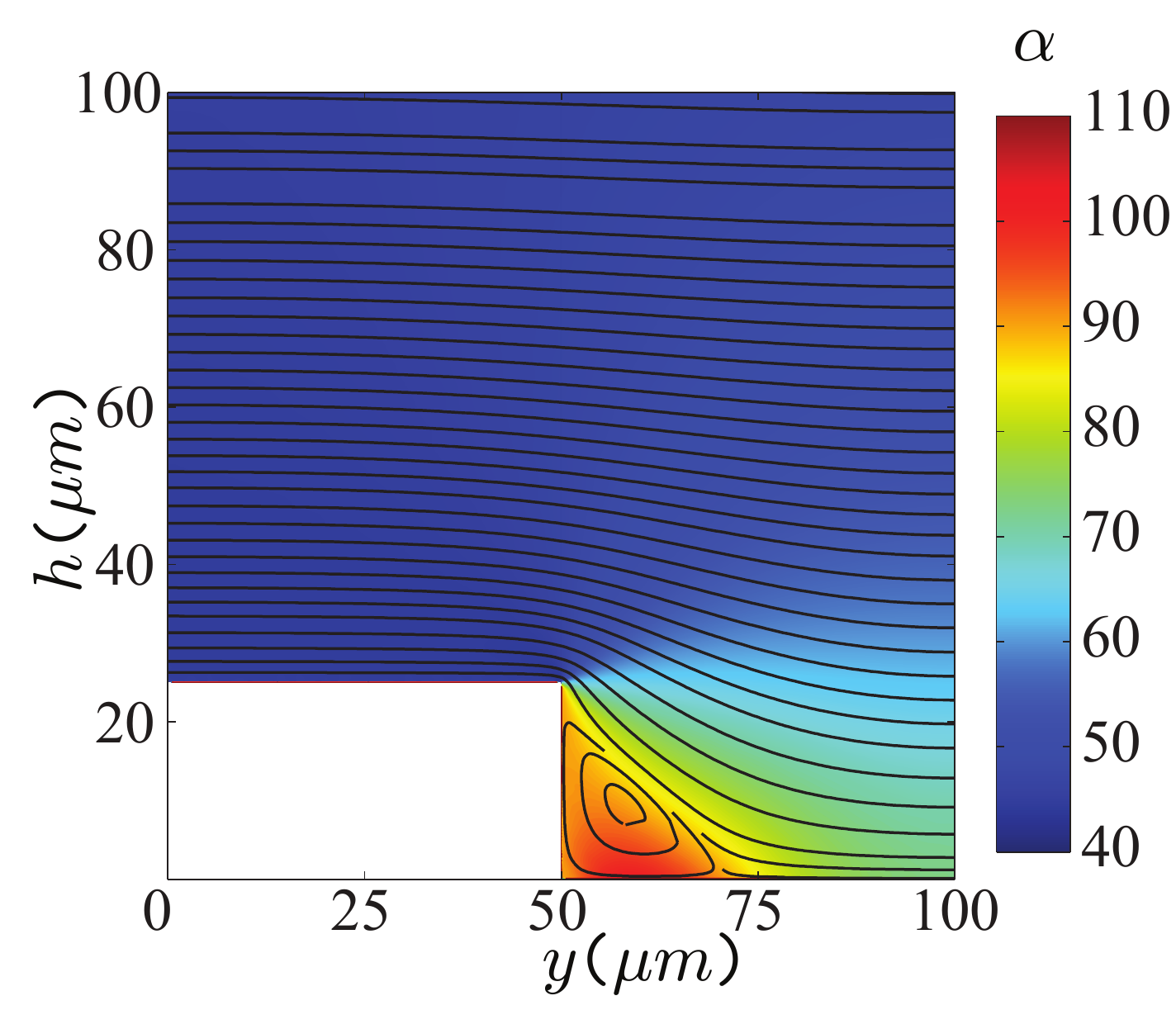}}\\ 
			 \end{tabular}
	\end{center}
\caption[Particle-free flow]{Numerical results showing the reorientation of the particle-free flow; $\alpha$ represents the angle that the projection of the velocity on the $x$--$y$ plane makes with respect to the $y$ axis. $(a)$ $\alpha$ as a function of the separation from the bottom wall of the cavities at different planes of constant $y$. The horizontal ($h = 24$  $\mu$m) and vertical ($\alpha = 45^{\circ}$) dashed lines mark the height of the ridges and the driving direction ($\Delta P_x = \Delta P_y$), respectively. The $y$ position corresponding to each curve is indicated in the inset with lines of the same style. $(b)$ Stream lines superposed on a contour plot of $\alpha$. The flow conforms with the imposed direction ($\alpha=45^{\circ}$) in the region above the ridges but deviates strongly inside the cavities, which guide flow along them ($\alpha \approx 90^{\circ}$)} 
\label{Fig:particle_free_flow}
\end{figure}

Insight into the motion of finite size particles can be gained by studying first the particle-free flow in the system, 
even though the exact trajectories of the particles cannot be predicted without considering hydrodynamic interactions and the effect 
of short-range particle-wall repulsive forces.\cite{Li:2007,Balvin:2009,Drazer:2009,Luo:2011,Bowman:2012}
For simplicity, we consider an infinite system without lateral confinement, i.e., no lateral walls in the main channel and therefore no recirculation above the ridges.
In the Stokes regime, we can decompose the flow field resulting from any orientation of the driving force into two independent components, corresponding to
the 2-D flows along and perpendicular to the ridges/cavities. The flow along the cavities ($x$ direction) is  unidirectional and the transverse flow ($y$ direction) 
corresponds to the well-studied flow over a rectangular cavity.\cite{Pan:1967,Higdon:1985,Shankar:1993,Meleshko:1996,Shankar:2000}
The combined 3-D flow then penetrates into the cavities to a different extent depending on their aspect ratio and also
exhibits recirculation regions close to the bottom corners of the cavities.\cite{Taneda:1979}
In order to obtain the local orientation of the velocity close to the patterned surface we computed the 3-D Stokes flow field numerically for the case
of a driving force oriented at $45^\circ$ (see ESI for details). 
We characterize the local direction of the flow by the angle $\alpha$ that the fluid velocity projected on the $x$--$y$ plane makes with respect to the $y$ direction, 
i.e.,  $\alpha = 90^\circ$ corresponds to flow along the cavity.  
Figure~\ref{Fig:particle_free_flow}$a$ shows $\alpha$ as a function of height at different cross-sections parallel to the ridges (planes of constant $y$).
The inset in Fig.~\ref{Fig:particle_free_flow}$a$ shows the direction of the velocity field in the vicinity of the patterned surface in a cross section perpendicular to the ridges.
The location of the cross-sections corresponding to  the different curves plotted in the Fig.~\ref{Fig:particle_free_flow}$a$  are indicated in the inset with lines of the same style; 
$y = $ 0 and 100 $\mu$m correspond to symmetry planes at the center of the ridge and cavity, respectively. 
Figure \ref{Fig:particle_free_flow}$a$ shows that the flow field is nearly parallel to the driving direction ($\alpha \approx 45^\circ$), everywhere except for the flow near or inside the shallow cavities 
created by the ridges. In fact, the cavities guide the flow along them, with $\alpha \approx 90^{\circ}$ near their center; values of $\alpha > 90^{\circ}$ indicate recirculation regions.
The streamlines, projected into a cross-section perpendicular to the ridges, are shown in Fig.~\ref{Fig:particle_free_flow}$b$. 
The fluid penetrates completely into the cavities and creates corotating recirculating regions near each corner. 
To consider the effect that the reorientation of the flow might have on different particles, the background in Fig.~\ref{Fig:particle_free_flow}$b$ is a contour plot for the velocity angle $\alpha$. 
It is clear that the deeper into the cavity a particle goes, the more it would be deflected by the flow along the ridges. The extent to which a particle enters the cavities depends, in general, on its settling velocity and size,
and two distinct regimes can be distinguished. In one case, if the particles are traversing the cavities much faster than their settling velocity, they can be separated based on their size,
with smaller particles typically entering deeper into the cavities and, as a result, exhibiting larger deflections. On the other hand, if the particle velocity across the cavities is comparable to the
settling velocity, particles of the same size can be separated by mass, with heavier particles showing a larger deflection along the cavities. 
In this latter case,  particles could remain constrained to flow along the cavities, if the flow that drags them out of the cavity on the reentrant corner 
is not sufficient to overcome their buoyant weight.

\section{Results and discussion}

\subsection{Separation of spherical particles}

\begin{figure}[htbp]
\begin{center}
\includegraphics[width=\columnwidth]{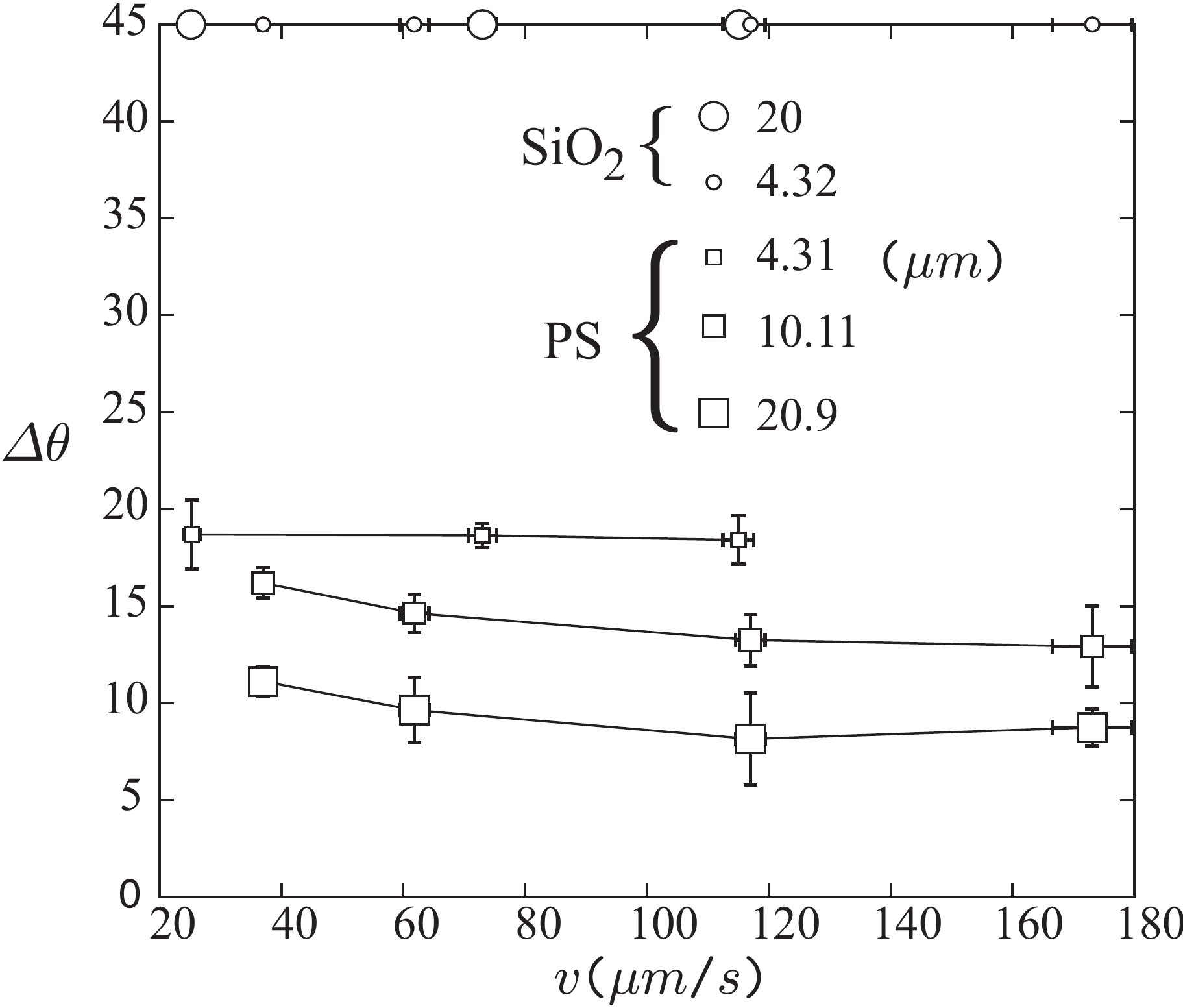} 
\caption{Deflection angle of silica (SiO$_2$) and polystyrene (PS) spherical particles of different size  in the device with large ridges ($h = 24$ $\mu$m, $w = 100$ $\mu$m) 
as a function of the approach velocity of the 4.32 $\mu$m SiO$_2$ particles in the plain region of the substrate before reaching the patterned area. 
The SiO$_2$ particles are confined to flow along the cavities over the range of velocities considered ($\alpha = 45^{\circ}$). 
PS particles can easily traverse the cavities and deflect to a less extent as their size increases for a given geometry of the ridges.}
\label{Fig:ParticlesLarge}
\end{center}
\end{figure}

First, we shall discuss the behavior of spherical particles of different size and density.
Figure~\ref{Fig:ParticlesLarge} shows the deflection angles 
obtained in the device with large ridges oriented at 45$^{\circ}$ with respect to the direction of the main channel (see Fig.~\ref{Fig:Device}).
As mentioned in section \ref{subsec:methods}, the particles are allowed to sediment to the bottom of the settling reservoir ($r$ in Fig. \ref{Fig:Device}$(a)$) before being flow focused towards the array of open cavities. The velocity of the 4.32 $\mu$m silica particles in the flat region of the device (before the slanted ridges) was used to characterize the flow velocity (see ESI for details).

\begin{figure}[htbp]
\begin{center}
\includegraphics[width=\columnwidth]{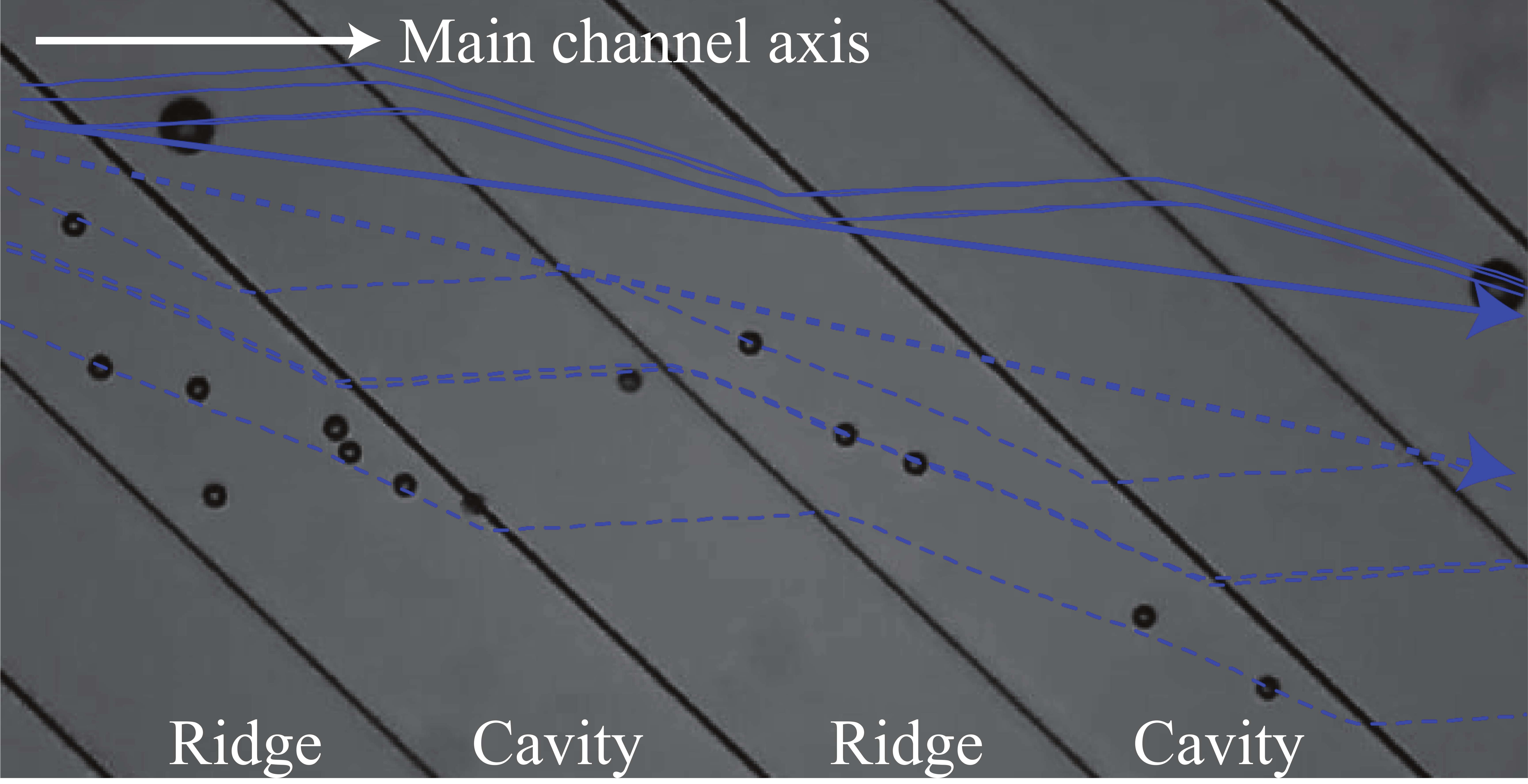}
\caption{Trajectories of the 10.11 (dotted lines) and 20.9 $\mu$m (solid lines) PS particles as they traverse the large ridges (the arrows represent mean migration directions). The particles move in the direction of the main flow when they are on top of the ridges but
are deflected as they traverse the cavities (see ESI).}
\label{Fig:Particles_tracks}
\end{center}
\end{figure}

First of all, we observe that SiO$_2$ particles (both 4.32 and 20 $\mu$m particles) are confined to move along the cavities 
($\Delta \theta = 45 ^{\circ}$), for all the flow velocities considered here.
In contrast, PS particles can easily move over the ridges and exhibit substantially smaller deflection angles compared to the SiO$_2$ particles (see ESI Movies 4 and 5). We note that, even at the largest flow rates, a few of the 10.11 $\mu$m PS particles become initially confined inside the cavities as they flow into the patterned region of the surface. This occurs for particles that circumvent, instead of going over, the ridges to go into the cavities, and could be avoided by flow focusing the particles more tightly to better control the entrance region. These particles remain confined during their motion through the observation window in the experiments and were not considered in the measurement of the deflection angle (see ESI Movie 4). 
Interestingly, the deflection angles for the different PS particles are nearly constant as a function of the flow velocity.
This indicates that particle sedimentation into the open cavities is not important for the range of velocities considered here.
More important, the deflection angles are clearly different for PS particles of different size, thus enabling their separation. 
Specifically, the average deflection of the PS particles decreases as the particle size increases, 
which suggests that their deflection is determined by the extent to which they are carried into the cavity by the flow,
with smaller particles reaching deeper into the cavities.
To demonstrate that the deflection of the particles originates in the flow along the cavities,  
in Fig.~\ref{Fig:Particles_tracks} we present the trajectories followed by different PS particles. 
It is clear that PS particles move in the direction of the main flow when they are on top of the ridges but are deflected as they traverse the cavities, 
which is consistent with the general features observed in the particle free flow field (see Fig.~\ref{Fig:particle_free_flow} and ESI). 

\begin{figure}[htbp]
\begin{center}
\includegraphics[width=\columnwidth]{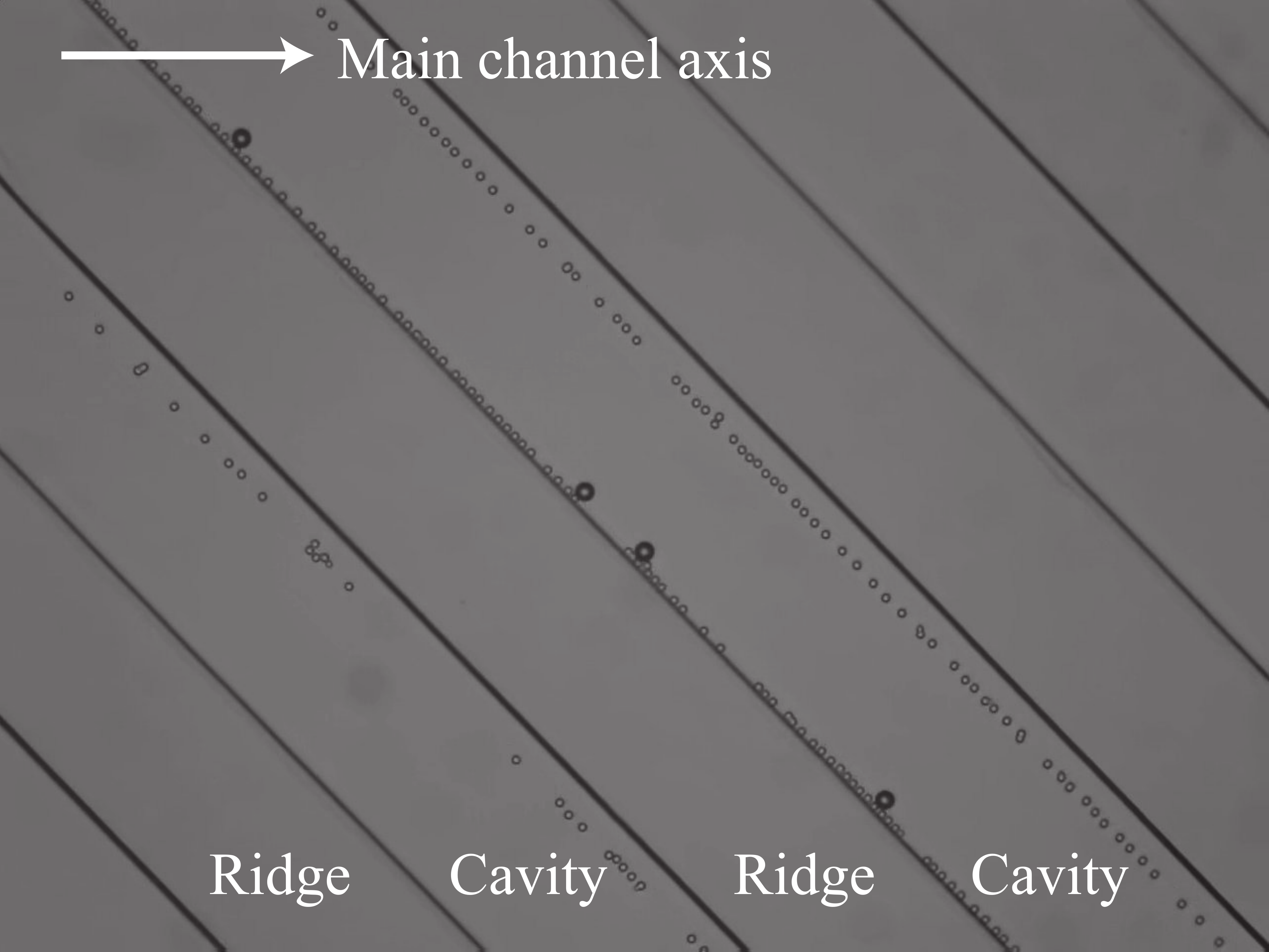}
\caption{Trajectories of the 4.32 $\mu$m silica particles confined to flow along the cavities in the device with large ridges. Particles moving in the vicinity of the downstream (or reentrant) corner attain an equilibrium position of a few microns away from ridge, while those moving along the upstream (or entrant) corner move in close proximity to the wall. This behavior is consistent with the presence and characteristic size of the corotating recirculating regions observed in the particle-free flow (see Fig.~\ref{Fig:particle_free_flow}$a$). A few of the 10 $\mu$m PS particles were also caught in the upstream recirculating region as they first entered the cavities (see ESI).}
\label{Fig:particle_tracks_confined}
\end{center}
\end{figure}

In these experiments, we observed that the small silica particles moving along the open cavities, and close to the bottom wall,
also migrate laterally to equilibrium positions on either corner of the cavities, as shown in Fig.~\ref{Fig:particle_tracks_confined} (see ESI).
Specifically, those particles moving in the vicinity of the downstream or reentrant corner attain an equilibrium position of a few microns away from ridge. 
On the other hand, the particles that are moving along the upstream (or entrant) corner move in close proximity to the edge of the ridge. 
These observation are consistent with the presence and characteristic size of the corotating recirculating regions 
that ensue in both corners of the cavities in the particle-free flow (see Fig.~\ref{Fig:particle_free_flow}$a$). 
The large differences in deflection angles shown in Figure~\ref{Fig:ParticlesLarge} indicate that very high purity separation is possible. We demonstrate the potential of this platform to fractionate a multicomponent sample with high purity of each individual component by separating a mixture of 4.32 SiO$_2$ particles, 10.11  $\mu$m PS particles, 
and 20.9 $\mu$m PS particles. The measure purity is larger than 99\% for the SiO$_2$ particles, 100\% for the 10.11  $\mu$m, and 94\% for the 20.9 $\mu$m PS particles(see ESI Movie 5 for details).

\begin{figure}[htbp]
\begin{center}
\includegraphics[width=\columnwidth]{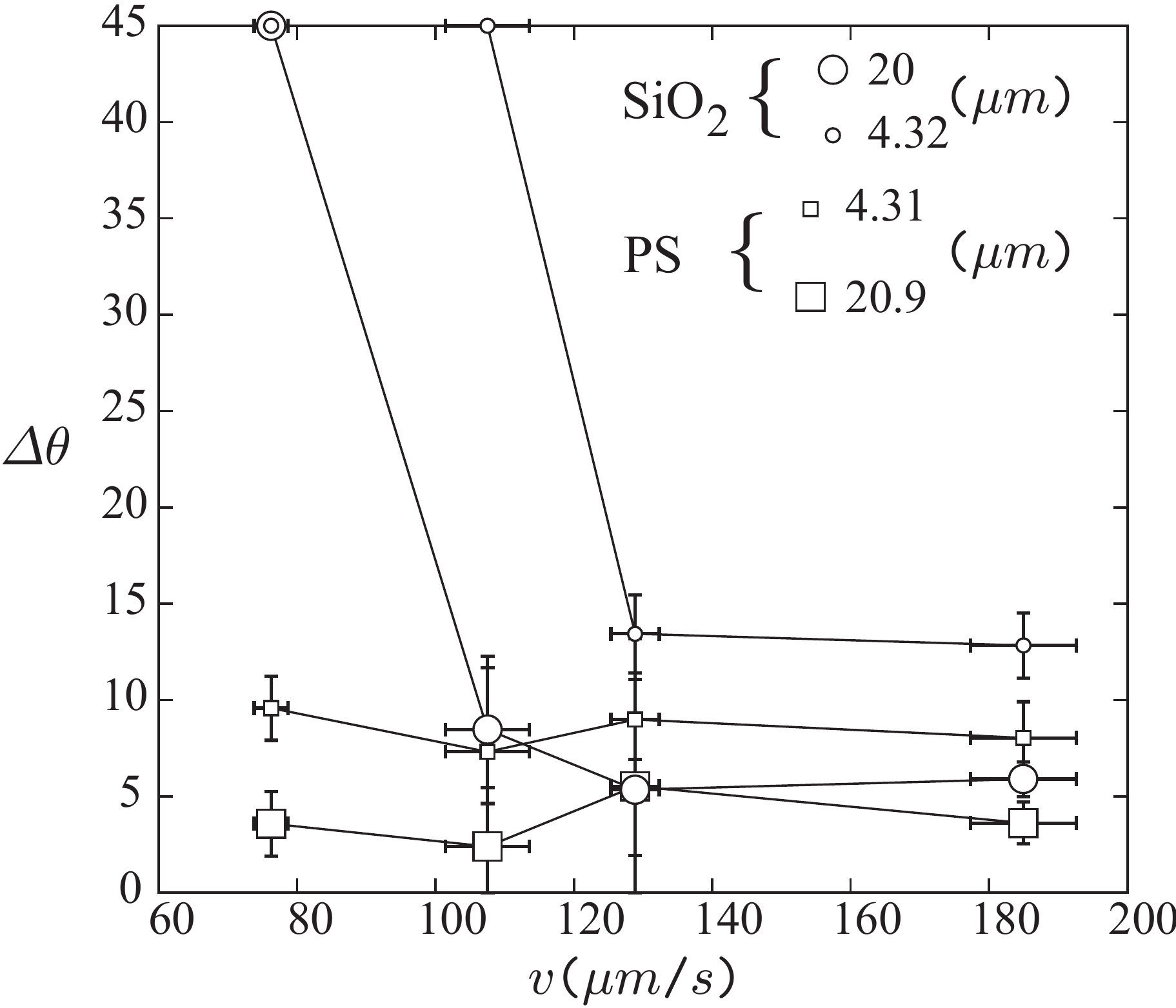}
\caption{Deflection angle of silica (SiO$_2$) and polystyrene (PS) spherical particles of different size in the device with small ridges ($h = 24$ $\mu$m, $w = 100$ $\mu$m) 
as a function of the approach velocity of the 4.32 $\mu$m SiO$_2$ particles in the plain region of the substrate before reaching the patterned area.}
\label{Fig:ParticlesSmall}
\end{center}
\end{figure}

Figure~\ref{Fig:ParticlesSmall} presents the deflection angles measured in the device with small ridges. 
The main difference with respect to the results discussed before is the fact that both sizes of SiO$_2$ particles are eventually able to move across the ridges as the flow rate increases. 
More important for separation purposes, there is a range of velocities for which the larger SiO$_2$ particles move across the ridges (with $\Delta \theta \lesssim 10^{\circ}$)
but the smaller SiO$_2$ particles remain confined and move along the cavities ($\Delta \theta = 45 ^{\circ}$). 
This is probably due to the larger drag force that the fluid exerts on the larger particles.  
Similarly to the case of the 10.11 $\mu$m PS particles in the device with large ridges, a few of the small silica particles can become trapped in the recirculating region at the corners of the cavities even at largest flow rate, and are thus not considered in the analysis. 
The PS particles exhibit analogous behavior to that observed in the experiments performed in the device with large ridges. 
It is clear that for particles with the same density, smaller particles deflect more and,
for particles of the same size, heavier ones deflect more.
The diminishing effect of sedimentation on the deflection angle as the flow rate increases, 
together with the increased effect of particle size at larger flow rates, 
are manifested by the cross-over that takes place in Fig. \ref{Fig:ParticlesSmall} for $v \sim110 \mu$m/s
between the deflection of small PS particles (4.31 $\mu$m) and large silica ones (20 $\mu$m).
At lower velocities, the density and sedimentation of the heavier particles dominates  and they deflect more than the PS particles, even in the absence of total confinement.
On the other hand, at higher velocities,  sedimentation into the cavities becomes negligible, and the small PS particles deflect more than the larger silica ones. 
 
 \subsection{Separation of blood cells}
\begin{figure}[htbp]
\includegraphics[width=\columnwidth]{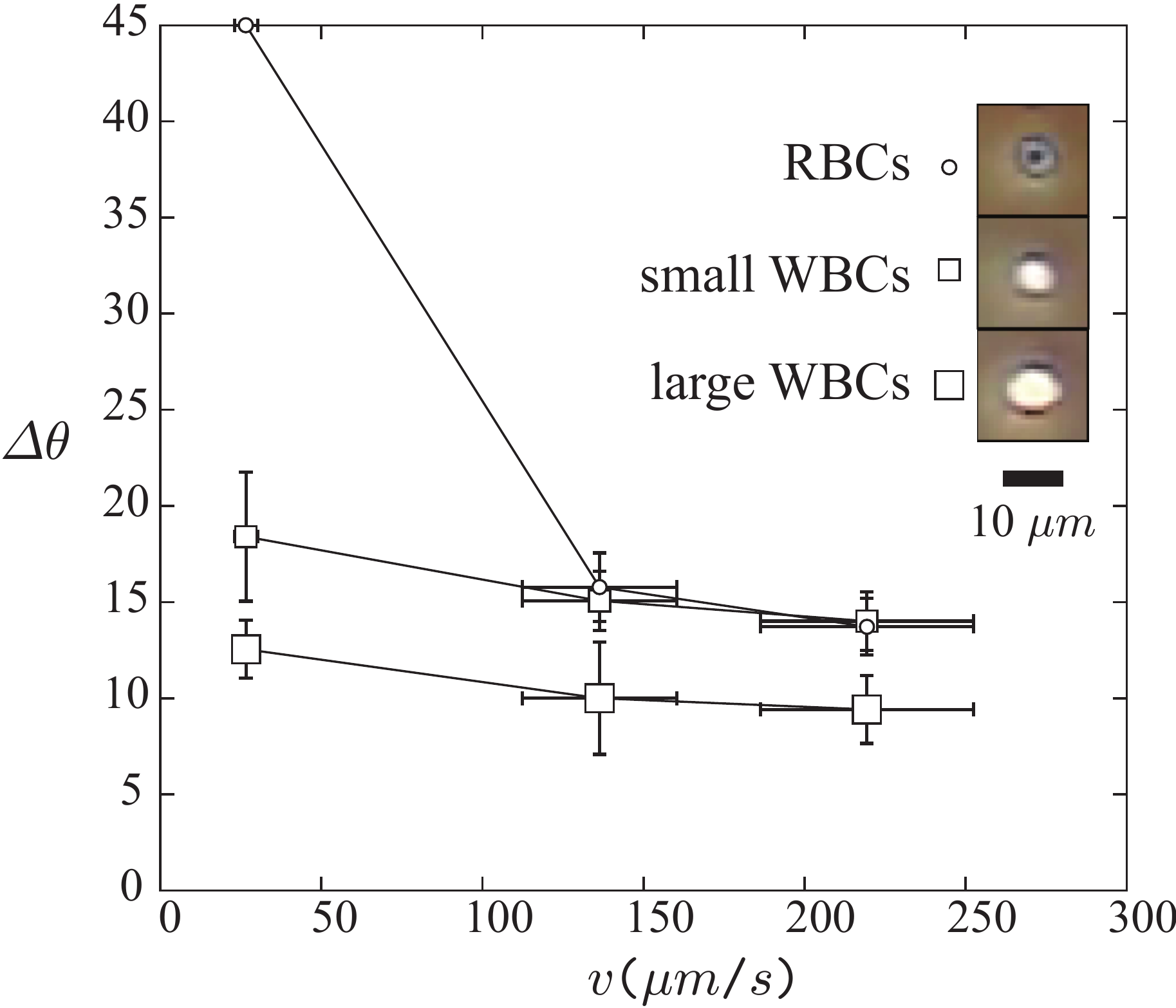}
\caption{Deflection angle of white and red blood cells (WBCs and RBCs) as a function of the approach velocity of the RBCs in the plain region of the device before the patterned area. Two subpopulations of WBCs could be clearly distinguished by size in the experiments.}
\label{Fig:blood.pdf}
\end{figure}

Next we used the device with large ridges to demonstrate the potential of this platform to separate different cell populations present in a blood sample 
based on cell size and density. RBCs are smaller and heavier, and should therefore deflect more than WBCs.
\cite{Pertoft:1980,Wheaters:2006} Platelets are the smallest cellular components in blood, with densities in the range spanned by that of WBCs.\cite{Corash:1977,Wheaters:2006,Inglis:2008}
Figure~\ref{Fig:blood.pdf} presents the deflection angle for RBCs and WBCs as a function of flow rate. 
In this case, we use the average velocity of RBCs in the flat region before they reach the patterned surface to characterize the flow velocity. 
It is clear that the dependence of the deflection angle of RBCs and WBCs on flow rate is qualitatively similar to the measurements performed with spherical particles, 
with the behavior of RBCs and WBCs resembling that of the silica and PS particles, respectively. 
Note, however, that the comparison with particle experiments is not quantitative, as expected based on the substantial difference in density between RBCs and silica particles.
RBCs deflect more than WBCs, and exhibit complete confinement inside the cavities at low flow rates.  
However, unlike the motion of silica particles observed in the present device,  RBCs can go over the ridges and move 
over the patterned surface as the flow rate increases (see ESI Movie 6). Again, this is consistent with our previous results given that RBCs are lighter than silica particles.
Two subpopulations of WBCs could be clearly distinguished in the experiments (see Materials and methods). 
Larger WBCs deflect to a lesser extent than smaller ones. 
As in the case of the PS particles, the deflection angle of WBCs does not change considerably over the range of velocities considered, 
which indicates that sedimentation does not significantly affect their motion. 
As mentioned before (see section \ref{subsec:methods}), 
platelets remain suspended across the height of the channel and we can take advantage of their small size and light buoyant weight  
to consider them as tracer particles to probe the flow. Their motion is entirely consistent with the characteristics of the particle free flow.
Specifically, platelets exhibited a broad range of deflection angles depending on their vertical position, with those moving far from the bottom surface not deflecting at all, and 
those moving close to the patterned surface showing the largest deflection angles (for simplicity, their deflection angle was not included in Fig.~\ref{Fig:blood.pdf} ).  
In fact, platelets were collected with 100\% purity in the outlet channel aligned with the injection channel, 
indicating that some of the platelets in the flow focused stream did not deflect at all. 
These observations also confirm that the effect of the patterned surface on the flow is localized to its vicinity and that the recirculation observed in confined geometries 
is negligible here.

\section{Conclusions}

We presented a microfluidic platform that allows the high-troughput, high-purity fractionation of suspensions based on the size and sedimentation velocity of the different components. 
Interestingly, the interplay between the sometimes competing effects of size and sedimentation velocity results in a highly versatile separation method.
The separation mechanism relies on the flow characteristics in the vicinity of slanted open cavities. 
The direction of the flow over the ridges creating the cavities does not deviate significantly from the imposed driving field. 
On the other hand, there is a significant change in flow direction in the vicinity and inside the open cavities, which essentially guide the flow along them. 
As a result, the deflection exhibited by different particles depends on the extent to which they penetrate into the open cavities. 
When sedimentation is negligible (light particles and high flow rates) 
smaller particles are advected by the fluid deeper into the open cavities,
 and therefore deflect more than bigger particles. 
We note that, in principle, remarkably high throughputs (with particle velocities of the order of 1 cm/s) could be achieved before inertial effects that might interfere with the underlying separation mechanism become important.\cite{Stone:2004} 
In the opposite limit (heavy particles and low flow rates), sedimentation becomes important and heavier particles deflect more.
In this regime, however, the maximum flow rate is limited by the settling velocity of the particles, which reduces the overall throughput of the system.
Other fields, however, e.g. electric, dielectrophoretic, and magnetic, can be easily integrated with the device to play or enhance the role of gravity,
which suggests that this technique could be readily adapted for a broad range of applications.

Born out of the above discussion, several applications are envisioned.  As an alternative to chemical lysis, this platform could be used as a microfluidic {\emph{centrifuge}}  because it allows the mild and non-destructive depletion and recovery of RBCs from the rest of the blood components. Furthermore, plasma proteins, bacterium, platelets, different subpopulations of WBCs, and other rare blood cells would migrate at different angles and can be continuously fractionated. More generally, this platform could be a non-clogging and contamination-free alternative for laboratories that frequently use filtration to separate cells of different size (from digested tissue, for example). Droplets of different size and density can also be separated, and this platform might also benefit droplet microfluidic applications.\cite{Teh:2008} The separation of cells, proteins, RNA and DNA, for example, can also be performed using centrifugal, electric and dielectrophoretic forces perpendicular to the patterned substrate to tune their settling velocity into the cavities and thus their deflection angle. In particular, patterning interdigitated electrodes between the ridges would be a simple way to integrate dielectrophoretic forces into the device. Target cells and biomolecules can be tagged with 
particles via specific marker-antibody interactions, and the properties of the tagging particles can be tailored to tune the settling velocity of the tagged species. Differences in settling velocity can also be used to control the time of arrival of different species to an array of open cavities. In this case, particles of the same or different size but with different settling velocity would eventually separate laterally after reaching the patterned surface at different times, even if sedimentation does not play a role in their migration angle. 
In summary, this is a promising microfluidic separation platform that could benefit a wide range of applications.

\section{Acknowledgements}
The authors would like to acknowledge Colin Paul and Matthew Dallas for insightful discussions and suggestions regarding the blood experiments. J. A. B. would like to acknowledge partial financial support from the INBT at Johns Hopkins University. 
This work was partially funded by a pilot grant from NIH U54CA143868 and partially supported by the National Science 
Foundation under Grant No. CBET-0954840.

\footnotesize{
\providecommand*{\mcitethebibliography}{\thebibliography}
\csname @ifundefined\endcsname{endmcitethebibliography}
{\let\endmcitethebibliography\endthebibliography}{}

}


\begin{mcitethebibliography}{53}
\providecommand*{\natexlab}[1]{#1}
\providecommand*{\mciteSetBstSublistMode}[1]{}
\providecommand*{\mciteSetBstMaxWidthForm}[2]{}
\providecommand*{\mciteBstWouldAddEndPuncttrue}
  {\def\EndOfBibitem{\unskip.}}
\providecommand*{\mciteBstWouldAddEndPunctfalse}
  {\let\EndOfBibitem\relax}
\providecommand*{\mciteSetBstMidEndSepPunct}[3]{}
\providecommand*{\mciteSetBstSublistLabelBeginEnd}[3]{}
\providecommand*{\EndOfBibitem}{}
\mciteSetBstSublistMode{f}
\mciteSetBstMaxWidthForm{subitem}
{(\emph{\alph{mcitesubitemcount}})}
\mciteSetBstSublistLabelBeginEnd{\mcitemaxwidthsubitemform\space}
{\relax}{\relax}

\bibitem[Blow(2007)]{Blow:2007}
N.~Blow, \emph{Nature Methods}, 2007, \textbf{4}, 665--670\relax
\mciteBstWouldAddEndPuncttrue
\mciteSetBstMidEndSepPunct{\mcitedefaultmidpunct}
{\mcitedefaultendpunct}{\mcitedefaultseppunct}\relax
\EndOfBibitem
\bibitem[Blow(2009)]{Blow:2009}
N.~Blow, \emph{Nature Methods}, 2009, \textbf{6}, 683--686\relax
\mciteBstWouldAddEndPuncttrue
\mciteSetBstMidEndSepPunct{\mcitedefaultmidpunct}
{\mcitedefaultendpunct}{\mcitedefaultseppunct}\relax
\EndOfBibitem
\bibitem[Pamme(2007)]{Pamme:2007}
N.~Pamme, \emph{Lab on a Chip}, 2007, \textbf{7}, 1644\relax
\mciteBstWouldAddEndPuncttrue
\mciteSetBstMidEndSepPunct{\mcitedefaultmidpunct}
{\mcitedefaultendpunct}{\mcitedefaultseppunct}\relax
\EndOfBibitem
\bibitem[{Kersaudy-Kerhoas} \emph{et~al.}(2008){Kersaudy-Kerhoas}, Dhariwal,
  and Desmulliez]{Kersaudy-Kerhoas:2008}
M.~{Kersaudy-Kerhoas}, R.~Dhariwal and M.~Desmulliez, \emph{{IET}
  Nanobiotechnology}, 2008, \textbf{2}, 1\relax
\mciteBstWouldAddEndPuncttrue
\mciteSetBstMidEndSepPunct{\mcitedefaultmidpunct}
{\mcitedefaultendpunct}{\mcitedefaultseppunct}\relax
\EndOfBibitem
\bibitem[Kulrattanarak \emph{et~al.}(2008)Kulrattanarak, van~der Sman,
  Schro{\"e}n, and Boom]{Kulrattanarak:2008}
T.~Kulrattanarak, R.~van~der Sman, C.~Schro{\"e}n and R.~Boom, \emph{Advances
  in Colloid and Interface Science}, 2008, \textbf{142}, 53--66\relax
\mciteBstWouldAddEndPuncttrue
\mciteSetBstMidEndSepPunct{\mcitedefaultmidpunct}
{\mcitedefaultendpunct}{\mcitedefaultseppunct}\relax
\EndOfBibitem
\bibitem[Lenshof and Laurell(2010)]{Lenshof:2010}
A.~Lenshof and T.~Laurell, \emph{Chemical Society Reviews}, 2010, \textbf{39},
  1203\relax
\mciteBstWouldAddEndPuncttrue
\mciteSetBstMidEndSepPunct{\mcitedefaultmidpunct}
{\mcitedefaultendpunct}{\mcitedefaultseppunct}\relax
\EndOfBibitem
\bibitem[Yamada \emph{et~al.}(2004)Yamada, Nakashima, and Seki]{Yamada:2004}
M.~Yamada, M.~Nakashima and M.~Seki, \emph{Anal. Chem.}, 2004, \textbf{76},
  5465--5471\relax
\mciteBstWouldAddEndPuncttrue
\mciteSetBstMidEndSepPunct{\mcitedefaultmidpunct}
{\mcitedefaultendpunct}{\mcitedefaultseppunct}\relax
\EndOfBibitem
\bibitem[Yamada and Seki(2005)]{Yamada:2005}
M.~Yamada and M.~Seki, \emph{Lab on a Chip}, 2005, \textbf{5}, 1233--1239\relax
\mciteBstWouldAddEndPuncttrue
\mciteSetBstMidEndSepPunct{\mcitedefaultmidpunct}
{\mcitedefaultendpunct}{\mcitedefaultseppunct}\relax
\EndOfBibitem
\bibitem[Di~Carlo \emph{et~al.}(2007)Di~Carlo, Irimia, Tompkins, and
  Toner]{DiCarlo:2007}
D.~Di~Carlo, D.~Irimia, R.~G. Tompkins and M.~Toner, \emph{Proceedings of the
  National Academy of Sciences of the United States of America}, 2007,
  \textbf{104}, 18892--18897\relax
\mciteBstWouldAddEndPuncttrue
\mciteSetBstMidEndSepPunct{\mcitedefaultmidpunct}
{\mcitedefaultendpunct}{\mcitedefaultseppunct}\relax
\EndOfBibitem
\bibitem[Di~Carlo \emph{et~al.}(2008)Di~Carlo, Edd, Irimia, Tompkins, and
  Toner]{DiCarlo:2008}
D.~Di~Carlo, J.~F. Edd, D.~Irimia, R.~G. Tompkins and M.~Toner, \emph{Anal.
  Chem.}, 2008, \textbf{80}, 2204--2211\relax
\mciteBstWouldAddEndPuncttrue
\mciteSetBstMidEndSepPunct{\mcitedefaultmidpunct}
{\mcitedefaultendpunct}{\mcitedefaultseppunct}\relax
\EndOfBibitem
\bibitem[Park \emph{et~al.}(2009)Park, Song, and Jung]{Park:2009}
J.~Park, S.~Song and H.~Jung, \emph{Lab on a Chip}, 2009, \textbf{9},
  939--948\relax
\mciteBstWouldAddEndPuncttrue
\mciteSetBstMidEndSepPunct{\mcitedefaultmidpunct}
{\mcitedefaultendpunct}{\mcitedefaultseppunct}\relax
\EndOfBibitem
\bibitem[Sim \emph{et~al.}(2011)Sim, Kwon, Park, Lee, and Jung]{Sim:2011}
T.~S. Sim, K.~Kwon, J.~C. Park, J.~Lee and H.~Jung, \emph{Lab on a Chip}, 2011,
  \textbf{11}, 93\relax
\mciteBstWouldAddEndPuncttrue
\mciteSetBstMidEndSepPunct{\mcitedefaultmidpunct}
{\mcitedefaultendpunct}{\mcitedefaultseppunct}\relax
\EndOfBibitem
\bibitem[Huang \emph{et~al.}(2004)Huang, Cox, Austin, and Sturm]{Huang:2004}
L.~R. Huang, E.~C. Cox, R.~H. Austin and J.~C. Sturm, \emph{Science}, 2004,
  \textbf{304}, 987--990\relax
\mciteBstWouldAddEndPuncttrue
\mciteSetBstMidEndSepPunct{\mcitedefaultmidpunct}
{\mcitedefaultendpunct}{\mcitedefaultseppunct}\relax
\EndOfBibitem
\bibitem[Inglis \emph{et~al.}(2006)Inglis, Davis, Austin, and
  Sturm]{Inglis:2006}
D.~W. Inglis, J.~A. Davis, R.~H. Austin and J.~C. Sturm, \emph{Lab on a Chip},
  2006, \textbf{6}, 655\relax
\mciteBstWouldAddEndPuncttrue
\mciteSetBstMidEndSepPunct{\mcitedefaultmidpunct}
{\mcitedefaultendpunct}{\mcitedefaultseppunct}\relax
\EndOfBibitem
\bibitem[Stroock \emph{et~al.}(2002)Stroock, Dertinger, Ajdari, Mezi{\'c},
  Stone, and Whitesides]{StroockScience:2002}
A.~D. Stroock, S.~K.~W. Dertinger, A.~Ajdari, I.~Mezi{\'c}, H.~A. Stone and
  G.~M. Whitesides, \emph{Science}, 2002, \textbf{295}, 647--651\relax
\mciteBstWouldAddEndPuncttrue
\mciteSetBstMidEndSepPunct{\mcitedefaultmidpunct}
{\mcitedefaultendpunct}{\mcitedefaultseppunct}\relax
\EndOfBibitem
\bibitem[Stroock \emph{et~al.}(2002)Stroock, Dertinger, Whitesides, and
  Ajdari]{StroockAC:2002}
A.~D. Stroock, S.~K. Dertinger, G.~M. Whitesides and A.~Ajdari, \emph{Anal.
  Chem.}, 2002, \textbf{74}, 5306--5312\relax
\mciteBstWouldAddEndPuncttrue
\mciteSetBstMidEndSepPunct{\mcitedefaultmidpunct}
{\mcitedefaultendpunct}{\mcitedefaultseppunct}\relax
\EndOfBibitem
\bibitem[Choi and Park(2007)]{Choi:2007}
S.~Choi and J.~Park, \emph{Lab on a Chip}, 2007, \textbf{7}, 890\relax
\mciteBstWouldAddEndPuncttrue
\mciteSetBstMidEndSepPunct{\mcitedefaultmidpunct}
{\mcitedefaultendpunct}{\mcitedefaultseppunct}\relax
\EndOfBibitem
\bibitem[Choi \emph{et~al.}(2008)Choi, Song, Choi, and Park]{Choi:2008}
S.~Choi, S.~Song, C.~Choi and J.~Park, \emph{Anal. Chem.}, 2008, \textbf{81},
  50--55\relax
\mciteBstWouldAddEndPuncttrue
\mciteSetBstMidEndSepPunct{\mcitedefaultmidpunct}
{\mcitedefaultendpunct}{\mcitedefaultseppunct}\relax
\EndOfBibitem
\bibitem[Chen and Gao(2008)]{Chen:2008}
H.~Chen and D.~Gao, \emph{Applied Physics Letters}, 2008, \textbf{92},
  173502--173502--3\relax
\mciteBstWouldAddEndPuncttrue
\mciteSetBstMidEndSepPunct{\mcitedefaultmidpunct}
{\mcitedefaultendpunct}{\mcitedefaultseppunct}\relax
\EndOfBibitem
\bibitem[Hsu \emph{et~al.}(2008)Hsu, Carlo, Chen, Irimia, and Toner]{Hsu:2008}
C.~Hsu, D.~D. Carlo, C.~Chen, D.~Irimia and M.~Toner, \emph{Lab Chip}, 2008,
  \textbf{8}, 2128--2134\relax
\mciteBstWouldAddEndPuncttrue
\mciteSetBstMidEndSepPunct{\mcitedefaultmidpunct}
{\mcitedefaultendpunct}{\mcitedefaultseppunct}\relax
\EndOfBibitem
\bibitem[K\u{r}iv\'{a}nkov\'{a} and Bo\u{c}ek(2005)]{Krivankova:2005}
L.~K\u{r}iv\'{a}nkov\'{a} and P.~Bo\u{c}ek, \emph{{ELECTROPHORESIS}}, 2005,
  \textbf{19}, 1064--1074\relax
\mciteBstWouldAddEndPuncttrue
\mciteSetBstMidEndSepPunct{\mcitedefaultmidpunct}
{\mcitedefaultendpunct}{\mcitedefaultseppunct}\relax
\EndOfBibitem
\bibitem[Rousselet \emph{et~al.}(1994)Rousselet, Salome, Ajdari, and
  Prostt]{Rousselet:1994}
J.~Rousselet, L.~Salome, A.~Ajdari and J.~Prostt, \emph{Nature}, 1994,
  \textbf{370}, 446--447\relax
\mciteBstWouldAddEndPuncttrue
\mciteSetBstMidEndSepPunct{\mcitedefaultmidpunct}
{\mcitedefaultendpunct}{\mcitedefaultseppunct}\relax
\EndOfBibitem
\bibitem[Kralj \emph{et~al.}(2006)Kralj, Lis, Schmidt, and Jensen]{Kralj:2006}
J.~G. Kralj, M.~T.~W. Lis, M.~A. Schmidt and K.~F. Jensen, \emph{Analytical
  Chemistry}, 2006, \textbf{78}, 5019--5025\relax
\mciteBstWouldAddEndPuncttrue
\mciteSetBstMidEndSepPunct{\mcitedefaultmidpunct}
{\mcitedefaultendpunct}{\mcitedefaultseppunct}\relax
\EndOfBibitem
\bibitem[Gl\"{u}ckstad(2004)]{Gluckstad:2004}
J.~Gl\"{u}ckstad, \emph{Nature Materials}, 2004, \textbf{3}, 9--10\relax
\mciteBstWouldAddEndPuncttrue
\mciteSetBstMidEndSepPunct{\mcitedefaultmidpunct}
{\mcitedefaultendpunct}{\mcitedefaultseppunct}\relax
\EndOfBibitem
\bibitem[Xiao and Grier(2010)]{Xiao:2010}
K.~Xiao and D.~G. Grier, \emph{Physical Review Letters}, 2010, \textbf{104},
  028302\relax
\mciteBstWouldAddEndPuncttrue
\mciteSetBstMidEndSepPunct{\mcitedefaultmidpunct}
{\mcitedefaultendpunct}{\mcitedefaultseppunct}\relax
\EndOfBibitem
\bibitem[Petersson \emph{et~al.}(2007)Petersson, \r{A}berg,
  {Sw\u{a}rd-Nilsson}, and Laurell]{Petersson:2007}
F.~Petersson, L.~\r{A}berg, A.~{Sw\u{a}rd-Nilsson} and T.~Laurell, \emph{Anal.
  Chem.}, 2007, \textbf{79}, 5117--5123\relax
\mciteBstWouldAddEndPuncttrue
\mciteSetBstMidEndSepPunct{\mcitedefaultmidpunct}
{\mcitedefaultendpunct}{\mcitedefaultseppunct}\relax
\EndOfBibitem
\bibitem[Inglis \emph{et~al.}(2004)Inglis, Riehn, Austin, and
  Sturm]{Inglis:2004}
D.~W. Inglis, R.~Riehn, R.~H. Austin and J.~C. Sturm, \emph{Applied Physics
  Letters}, 2004, \textbf{85}, 5093\relax
\mciteBstWouldAddEndPuncttrue
\mciteSetBstMidEndSepPunct{\mcitedefaultmidpunct}
{\mcitedefaultendpunct}{\mcitedefaultseppunct}\relax
\EndOfBibitem
\bibitem[Pamme and Wilhelm(2006)]{Pamme:2006}
N.~Pamme and C.~Wilhelm, \emph{Lab on a Chip}, 2006, \textbf{6}, 974\relax
\mciteBstWouldAddEndPuncttrue
\mciteSetBstMidEndSepPunct{\mcitedefaultmidpunct}
{\mcitedefaultendpunct}{\mcitedefaultseppunct}\relax
\EndOfBibitem
\bibitem[Liu \emph{et~al.}(2009)Liu, Stakenborg, Peeters, and Lagae]{Liu:2009}
C.~Liu, T.~Stakenborg, S.~Peeters and L.~Lagae, \emph{Journal of Applied
  Physics}, 2009, \textbf{105}, 102014\relax
\mciteBstWouldAddEndPuncttrue
\mciteSetBstMidEndSepPunct{\mcitedefaultmidpunct}
{\mcitedefaultendpunct}{\mcitedefaultseppunct}\relax
\EndOfBibitem
\bibitem[Giddings(1984)]{Giddings:1984}
J.~C. Giddings, \emph{Anal. Chem.}, 1984, \textbf{56}, 1258A--1270A\relax
\mciteBstWouldAddEndPuncttrue
\mciteSetBstMidEndSepPunct{\mcitedefaultmidpunct}
{\mcitedefaultendpunct}{\mcitedefaultseppunct}\relax
\EndOfBibitem
\bibitem[Tia and Herr(2009)]{Tia:2009}
S.~Tia and A.~E. Herr, \emph{Lab on a Chip}, 2009, \textbf{9}, 2524\relax
\mciteBstWouldAddEndPuncttrue
\mciteSetBstMidEndSepPunct{\mcitedefaultmidpunct}
{\mcitedefaultendpunct}{\mcitedefaultseppunct}\relax
\EndOfBibitem
\bibitem[Dorfman and Brenner(2001)]{Dorfman:2001}
K.~D. Dorfman and H.~Brenner, \emph{Journal of Colloid and Interface Science},
  2001, \textbf{238}, 390--413\relax
\mciteBstWouldAddEndPuncttrue
\mciteSetBstMidEndSepPunct{\mcitedefaultmidpunct}
{\mcitedefaultendpunct}{\mcitedefaultseppunct}\relax
\EndOfBibitem
\bibitem[Arata and Alexeev(2009)]{Arata:2009}
J.~P. Arata and A.~Alexeev, \emph{Soft Matter}, 2009, \textbf{5},
  2721--2724\relax
\mciteBstWouldAddEndPuncttrue
\mciteSetBstMidEndSepPunct{\mcitedefaultmidpunct}
{\mcitedefaultendpunct}{\mcitedefaultseppunct}\relax
\EndOfBibitem
\bibitem[Mao and Alexeev(2011)]{Mao:2011}
W.~Mao and A.~Alexeev, \emph{Physics of Fluids}, 2011, \textbf{23},
  051704--051704--4\relax
\mciteBstWouldAddEndPuncttrue
\mciteSetBstMidEndSepPunct{\mcitedefaultmidpunct}
{\mcitedefaultendpunct}{\mcitedefaultseppunct}\relax
\EndOfBibitem
\bibitem[Bernate and Drazer(2011)]{Bernate:2011}
J.~A. Bernate and G.~Drazer, \emph{Journal of Colloid and Interface Science},
  2011, \textbf{356}, 341--351\relax
\mciteBstWouldAddEndPuncttrue
\mciteSetBstMidEndSepPunct{\mcitedefaultmidpunct}
{\mcitedefaultendpunct}{\mcitedefaultseppunct}\relax
\EndOfBibitem
\bibitem[Bernate and Drazer(2012)]{Bernate:2012}
J.~A. Bernate and G.~Drazer, \emph{Physical Review Letters}, 2012,
  \textbf{108}, 214501\relax
\mciteBstWouldAddEndPuncttrue
\mciteSetBstMidEndSepPunct{\mcitedefaultmidpunct}
{\mcitedefaultendpunct}{\mcitedefaultseppunct}\relax
\EndOfBibitem
\bibitem[Li and Drazer(2007)]{Li:2007}
Z.~Li and G.~Drazer, \emph{Phys. Rev. Lett.}, 2007, \textbf{98}, 050602\relax
\mciteBstWouldAddEndPuncttrue
\mciteSetBstMidEndSepPunct{\mcitedefaultmidpunct}
{\mcitedefaultendpunct}{\mcitedefaultseppunct}\relax
\EndOfBibitem
\bibitem[Balvin \emph{et~al.}(2009)Balvin, Sohn, Iracki, Drazer, and
  Frechette]{Balvin:2009}
M.~Balvin, E.~Sohn, T.~Iracki, G.~Drazer and J.~Frechette, \emph{Physical
  Review Letters}, 2009, \textbf{103}, 078301--1 -- 078301--4\relax
\mciteBstWouldAddEndPuncttrue
\mciteSetBstMidEndSepPunct{\mcitedefaultmidpunct}
{\mcitedefaultendpunct}{\mcitedefaultseppunct}\relax
\EndOfBibitem
\bibitem[Frechette and Drazer(2009)]{Drazer:2009}
J.~Frechette and G.~Drazer, \emph{Journal of Fluid Mechanics}, 2009,
  \textbf{627}, 379--401\relax
\mciteBstWouldAddEndPuncttrue
\mciteSetBstMidEndSepPunct{\mcitedefaultmidpunct}
{\mcitedefaultendpunct}{\mcitedefaultseppunct}\relax
\EndOfBibitem
\bibitem[Luo \emph{et~al.}(2011)Luo, Sweeney, Risbud, Drazer, and
  Frechette]{Luo:2011}
M.~Luo, F.~Sweeney, S.~R. Risbud, G.~Drazer and J.~Frechette, \emph{Applied
  Physics Letters}, 2011, \textbf{99}, 064102--064102--3\relax
\mciteBstWouldAddEndPuncttrue
\mciteSetBstMidEndSepPunct{\mcitedefaultmidpunct}
{\mcitedefaultendpunct}{\mcitedefaultseppunct}\relax
\EndOfBibitem
\bibitem[Bowman \emph{et~al.}(2012)Bowman, Frechette, and Drazer]{Bowman:2012}
T.~Bowman, J.~Frechette and G.~Drazer, \emph{Lab Chip}, 2012, \textbf{12},
  2903--2908\relax
\mciteBstWouldAddEndPuncttrue
\mciteSetBstMidEndSepPunct{\mcitedefaultmidpunct}
{\mcitedefaultendpunct}{\mcitedefaultseppunct}\relax
\EndOfBibitem
\bibitem[Pan and Acrivos(1967)]{Pan:1967}
F.~Pan and A.~Acrivos, \emph{Journal of Fluid Mechanics}, 1967, \textbf{28},
  643--655\relax
\mciteBstWouldAddEndPuncttrue
\mciteSetBstMidEndSepPunct{\mcitedefaultmidpunct}
{\mcitedefaultendpunct}{\mcitedefaultseppunct}\relax
\EndOfBibitem
\bibitem[Higdon(1985)]{Higdon:1985}
J.~J.~L. Higdon, \emph{Journal of Fluid Mechanics}, 1985, \textbf{159},
  195--226\relax
\mciteBstWouldAddEndPuncttrue
\mciteSetBstMidEndSepPunct{\mcitedefaultmidpunct}
{\mcitedefaultendpunct}{\mcitedefaultseppunct}\relax
\EndOfBibitem
\bibitem[Shankar(1993)]{Shankar:1993}
P.~N. Shankar, \emph{Journal of Fluid Mechanics}, 1993, \textbf{250},
  371--383\relax
\mciteBstWouldAddEndPuncttrue
\mciteSetBstMidEndSepPunct{\mcitedefaultmidpunct}
{\mcitedefaultendpunct}{\mcitedefaultseppunct}\relax
\EndOfBibitem
\bibitem[Meleshko(1996)]{Meleshko:1996}
V.~V. Meleshko, \emph{Proceedings of the Royal Society A: Mathematical,
  Physical and Engineering Sciences}, 1996, \textbf{452}, 1999--2022\relax
\mciteBstWouldAddEndPuncttrue
\mciteSetBstMidEndSepPunct{\mcitedefaultmidpunct}
{\mcitedefaultendpunct}{\mcitedefaultseppunct}\relax
\EndOfBibitem
\bibitem[Shankar and Deshpande(2000)]{Shankar:2000}
P.~N. Shankar and M.~D. Deshpande, \emph{Annual Review of Fluid Mechanics},
  2000, \textbf{32}, 93--136\relax
\mciteBstWouldAddEndPuncttrue
\mciteSetBstMidEndSepPunct{\mcitedefaultmidpunct}
{\mcitedefaultendpunct}{\mcitedefaultseppunct}\relax
\EndOfBibitem
\bibitem[Taneda(1979)]{Taneda:1979}
S.~Taneda, \emph{Journal of the Physical Society of Japan}, 1979, \textbf{46},
  1935--1942\relax
\mciteBstWouldAddEndPuncttrue
\mciteSetBstMidEndSepPunct{\mcitedefaultmidpunct}
{\mcitedefaultendpunct}{\mcitedefaultseppunct}\relax
\EndOfBibitem
\bibitem[Pertoft \emph{et~al.}(1980)Pertoft, Johnsson, W\"armeg\r{a}rd, and
  Seljelid]{Pertoft:1980}
H.~Pertoft, A.~Johnsson, B.~W\"armeg\r{a}rd and R.~Seljelid, \emph{Journal of
  Immunological Methods}, 1980, \textbf{33}, 221--229\relax
\mciteBstWouldAddEndPuncttrue
\mciteSetBstMidEndSepPunct{\mcitedefaultmidpunct}
{\mcitedefaultendpunct}{\mcitedefaultseppunct}\relax
\EndOfBibitem
\bibitem[Stevens \emph{et~al.}(2006)Stevens, Young, Lowe, Deakin, and
  Heath]{Wheaters:2006}
A.~Stevens, B.~Young, J.~S. Lowe, P.~J. Deakin and J.~W. Heath, \emph{Wheater's
  Functional Histology: A Text and Colour Atlas}, Churchill Livingstone, 5th
  edn, 2006\relax
\mciteBstWouldAddEndPuncttrue
\mciteSetBstMidEndSepPunct{\mcitedefaultmidpunct}
{\mcitedefaultendpunct}{\mcitedefaultseppunct}\relax
\EndOfBibitem
\bibitem[Corash \emph{et~al.}(1977)Corash, Tan, and Gralnick]{Corash:1977}
L.~Corash, H.~Tan and H.~R. Gralnick, \emph{Blood}, 1977, \textbf{49},
  71--87\relax
\mciteBstWouldAddEndPuncttrue
\mciteSetBstMidEndSepPunct{\mcitedefaultmidpunct}
{\mcitedefaultendpunct}{\mcitedefaultseppunct}\relax
\EndOfBibitem
\bibitem[Inglis \emph{et~al.}(2008)Inglis, Morton, Davis, Zieziulewicz,
  Lawrence, Austin, and Sturm]{Inglis:2008}
D.~W. Inglis, K.~J. Morton, J.~A. Davis, T.~J. Zieziulewicz, D.~A. Lawrence,
  R.~H. Austin and J.~C. Sturm, \emph{Lab Chip}, 2008, \textbf{8},
  925--931\relax
\mciteBstWouldAddEndPuncttrue
\mciteSetBstMidEndSepPunct{\mcitedefaultmidpunct}
{\mcitedefaultendpunct}{\mcitedefaultseppunct}\relax
\EndOfBibitem
\bibitem[Stone \emph{et~al.}(2004)Stone, Stroock, and Ajdari]{Stone:2004}
H.~Stone, A.~Stroock and A.~Ajdari, \emph{Annual Review of Fluid Mechanics},
  2004, \textbf{36}, 381--411\relax
\mciteBstWouldAddEndPuncttrue
\mciteSetBstMidEndSepPunct{\mcitedefaultmidpunct}
{\mcitedefaultendpunct}{\mcitedefaultseppunct}\relax
\EndOfBibitem
\bibitem[Teh \emph{et~al.}(2008)Teh, Lin, Hung, and Lee]{Teh:2008}
S.~Teh, R.~Lin, L.~Hung and A.~P. Lee, \emph{Lab on a Chip}, 2008, \textbf{8},
  198\relax
\mciteBstWouldAddEndPuncttrue
\mciteSetBstMidEndSepPunct{\mcitedefaultmidpunct}
{\mcitedefaultendpunct}{\mcitedefaultseppunct}\relax
\EndOfBibitem
\end{mcitethebibliography}
\end{document}